\documentclass[twocolumn,prd,superscriptaddress,altaffilletter,nofootinbib]{revtex4-1}
\usepackage{float}
\usepackage{amsmath}
\usepackage{amssymb}
\usepackage{amsfonts}
\usepackage{comment}
\usepackage{graphicx}
\usepackage{hyperref}
\usepackage[nolist]{acronym}
\usepackage[usenames,dvipsnames]{xcolor}
\usepackage{xspace}
\usepackage{tikz}
\usepackage[squaren]{SIunits}
\usetikzlibrary{arrows,shapes,trees,decorations.pathreplacing}
\allowdisplaybreaks

\newcommand{\likelihood}{\mathcal{L}}

\newcommand{\Cov}{\pmb{\Sigma}}
\newcommand{\DDeff}{\vec{\Delta \ln D_\text{eff}}}
\newcommand{\Extmod}{\vec{\lambda}_\mathrm{mi}}

\newcommand{\DX}{\vec{\Delta x}}

\newcommand{\Ext}{\vec{\lambda}}
\newcommand{\Dext}{\vec{\Delta\lambda}}
\newcommand{\Deff}{\vec{D_\text{eff}}}
\newcommand{\Dh}{\vec{D_H}}
\newcommand{\Ob}{\vec{O}}
\newcommand{\SNR}{\vec{\rho}}
\newcommand{\CHISQ}{\vec{\xi^2}}
\newcommand{\PHI}{\vec{\phi}}
\newcommand{\T}{\vec{t}}

\newcommand{\DPHI}{\vec{\Delta\phi}}
\newcommand{\DT}{\vec{\Delta t}}
\newcommand{\prob}[1]{P \left( #1 \right) } 
\newcommand{\cprob}[2]{P (\, #1 \mid #2 \,) } 
\newcommand{\sh}{\mathrm{s}} 
\newcommand{\nh}{\mathrm{n}} 
\newcommand{\magnitude}[1]{\left|\, #1 \,\right|}

\newcommand{\gstlal}{GstLAL\xspace}

\newcommand{\hplus}{\ensuremath{h_{+}}\xspace}

\newcommand{\hcross}{\ensuremath{h_{\times}}\xspace}

\begin{document}

\title{Fast evaluation of multi-detector consistency for real-time gravitational wave searches}

\author{Chad Hanna}
\affiliation{Department of Physics, The Pennsylvania State University, University Park, PA 16802, USA}
\affiliation{Department of Astronomy and Astrophysics, The Pennsylvania State University, University Park, PA 16802, USA}
\affiliation{Institute for Gravitation and the Cosmos, The Pennsylvania State University, University Park, PA 16802, USA}
\affiliation{Institute for CyberScience, The Pennsylvania State University, University Park, PA 16802, USA}

\author{Sarah Caudill}
\affiliation{Leonard E.\ Parker Center for Gravitation, Cosmology, and Astrophysics, University of Wisconsin-Milwaukee, Milwaukee, WI 53201, USA}
\affiliation{Nikhef, Science Park, 1098 XG Amsterdam, Netherlands}

\author{Cody Messick}
\affiliation{Department of Physics, The Pennsylvania State University, University Park, PA 16802, USA}
\affiliation{Institute for Gravitation and the Cosmos, The Pennsylvania State University, University Park, PA 16802, USA}

\author{Amit Reza}
\affiliation{Department of Physics, Indian Institute of Technology Gandhinagar, Gujarat 382355, India}

\author{Surabhi Sachdev}
\affiliation{Department of Physics, The Pennsylvania State University, University Park, PA 16802, USA}
\affiliation{Institute for Gravitation and the Cosmos, The Pennsylvania State University, University Park, PA 16802, USA}
\affiliation{LIGO Laboratory, California Institute of Technology, MS 100-36, Pasadena, California 91125, USA}

\author{Leo Tsukada}
\affiliation{RESCEU, The University of Tokyo, Tokyo, 113-0033, Japan}
\affiliation{Graduate School of Science, The University of Tokyo, Tokyo 113-0033, Japan}

\author{Kipp Cannon}
\affiliation{Canadian Institute for Theoretical Astrophysics, 60 St. George Street, University of Toronto, Toronto, Ontario, M5S 3H8, Canada}
\affiliation{RESCEU, The University of Tokyo, Tokyo, 113-0033, Japan}

\author{Kent Blackburn}
\affiliation{LIGO Laboratory, California Institute of Technology, MS 100-36, Pasadena, California 91125, USA}

\author{Jolien D. E. Creighton}
\affiliation{Leonard E.\ Parker Center for Gravitation, Cosmology, and Astrophysics, University of Wisconsin-Milwaukee, Milwaukee, WI 53201, USA}

\author{Heather Fong}
\affiliation{Canadian Institute for Theoretical Astrophysics, 60 St. George Street, University of Toronto, Toronto, Ontario, M5S 3H8, Canada}

\author{Patrick Godwin}
\affiliation{Department of Physics, The Pennsylvania State University, University Park, PA 16802, USA}
\affiliation{Institute for Gravitation and the Cosmos, The Pennsylvania State University, University Park, PA 16802, USA}

\author{Shasvath Kapadia}
\affiliation{Leonard E.\ Parker Center for Gravitation, Cosmology, and Astrophysics, University of Wisconsin-Milwaukee, Milwaukee, WI 53201, USA}

\author{Tjonnie G. F. Li}
\affiliation{Department of Physics, The Chinese University of Hong Kong, Shatin, New Territories, Hong Kong}

\author{Ryan Magee}
\affiliation{Department of Physics, The Pennsylvania State University, University Park, PA 16802, USA}
\affiliation{Institute for Gravitation and the Cosmos, The Pennsylvania State University, University Park, PA 16802, USA}

\author{Duncan Meacher}
\affiliation{Department of Physics, The Pennsylvania State University, University Park, PA 16802, USA}
\affiliation{Institute for Gravitation and the Cosmos, The Pennsylvania State University, University Park, PA 16802, USA}

\author{Debnandini Mukherjee}
\affiliation{Leonard E.\ Parker Center for Gravitation, Cosmology, and Astrophysics, University of Wisconsin-Milwaukee, Milwaukee, WI 53201, USA}

\author{Alex Pace}
\affiliation{Department of Physics, The Pennsylvania State University, University Park, PA 16802, USA}
\affiliation{Institute for Gravitation and the Cosmos, The Pennsylvania State University, University Park, PA 16802, USA}

\author{Stephen Privitera}
\affiliation{Albert-Einstein-Institut, Max-Planck-Institut f{\"u}r Gravitationsphysik, D-14476 Potsdam-Golm, Germany}

\author{Rico K. L. Lo}
\affiliation{Department of Physics, The Chinese University of Hong Kong, Shatin, New Territories, Hong Kong}
\affiliation{LIGO Laboratory, California Institute of Technology, MS 100-36, Pasadena, California 91125, USA}

\author{Leslie Wade}
\affiliation{Department of Physics, Hayes Hall, Kenyon College, Gambier, Ohio 43022, USA}

\date{\today}
\begin{abstract}
Gravitational waves searches for compact binary mergers with LIGO and Virgo are
presently a two stage process. First, a gravitational wave signal is
identified. Then, an exhaustive search over possible signal parameters is
performed.  It is critical that the identification stage is efficient in order
to maximize the number of gravitational wave sources that are identified.
Initial identification of gravitational wave signals with LIGO and Virgo
happens in real-time which requires that less than one second of computational
time must be used for each one second of gravitational wave data collected. In
contrast, subsequent parameter estimation may require hundreds of hours of
computational time to analyze the same one second of gravitational wave data.
The real-time identification requirement necessitates efficient and often
approximate methods for signal analysis.  We describe one piece of real-time
gravitational-wave identification: an efficient method for ascertaining a
signal's consistency between multiple gravitational wave detectors suitable for
real-time gravitational wave searches for compact binary mergers.  This
technique was used in analyses of Advanced LIGO's second observing run and
Advanced Virgo's first observing run.
\end{abstract}

\maketitle

\section{Introduction}\label{sec:intro}

Advanced LIGO~\cite{harry2010advanced, aasi2015advanced} and
Virgo~\cite{acernese2014advanced} gravitational-wave observatories have seen
great success since the discovery of gravitational waves in September
2015~\cite{abbott2016observation}.  With the detection of gravitational waves
from ten binary black holes and one binary neutron star in just 170 days of
observing, gravitational wave detections are now happening at a rate of about
two per month of multi-detector observation time~\cite{ligo2018gwtc}. The rate
is expected to increase by more than an order of magnitude as the world-wide
network of gravitational wave detectors improve in
sensitivity~\cite{aasi2016prospects}.  Furthermore, over the next decade, the
gravitational wave detector network will grow to include
KAGRA~\cite{somiya2012detector, aso2013interferometer} in Japan and
LIGO-India~\cite{unnikrishnan2013indigo}.  It is anticipated that real-time
gravitational wave searches will use data from the entire gravitational wave
detector network in order to increase the detection rate and better localize
the gravitational wave sources~\cite{aasi2016prospects}.

The current generation of interferometric gravitational wave detectors measure
only one of two possible gravitational wave polarizations \hplus and \hcross.
However, given the detector orientation with respect to incoming signals, they
each measure a different linear combination of polarizations defined by a
suitable Earth-centered coordinate system. Furthermore, the response to the
gravitational wave signal changes as a function of the incident angle.  While
the detector response function is broad, some incident angles produce no
response in a given gravitational wave detector.  Thus, each detector's
response to a given signal is a function of the location and polarization state
of a gravitational wave source~\cite{abbott2009ligo, harry2010advanced,
acernese2014advanced}.  Given a particular gravitational wave signal incident
on Earth, it is possible to predict precisely what the projection of the signal
will be on each of the gravitational wave detectors~\cite{andersonbeam}.  A
corollary of this is that only certain arrival times, phases and amplitudes
measured in a set of gravitational wave detectors are consistent with a real
gravitational wave signal.  

Typically, gravitational wave searches have relied on one of several mechanisms
for imposing consistency of gravitational wave detection among a network of
detectors. Each method has a varying degree of fidelity traded against
complexity or computational cost.  The ideal method requires explicitly
evaluating the likelihood of measuring the projected waveform in each detector
while sampling over physically reasonable prior distributions, e.g.,
isotropically distributed on the sky, using Markov Chain Monte Carlo
techniques~\cite{veitch2015parameter}. This is the method used in parameter
estimation as the second stage of gravitational wave astronomy.  While regarded
as the most accurate assessment of consistency among gravitational wave
detectors, the full likelihood evaluation has the practical limitation of being
a computationally costly procedure.  However, approximate methods have been
developed~\cite{singer2016rapid}, that are used to e.g., quickly compute the
sky location of gravitational wave events.  The approximate methods are orders
of magnitude faster and still extremely accurate. Nevertheless, the approximate
methods used to compute sky maps are not suitably fast to use for the initial
identification stage of gravitational wave events.  

Another widely adopted mechanism to ensure consistency between gravitational
wave observatories is to form a coherent combination of the detector
data~\cite{pai2002computational,macleod2016fully} in order to create a sky
location dependent reconstruction of the two gravitational wave polarizations.
This technique typically uses a predefined grid of sky positions.  Coherence
is used presently in searches for gravitational wave
bursts~\cite{klimenko2008coherent, sutton2010x} and triggered searches for
gamma ray bursts~\cite{harry2011targeted, williamson2014improved} which only
focus on specific time intervals and sky regions that overlap with known gamma
ray bursts.

The final mechanism involves searching each detector data stream independently,
triggering on potential signal times by finding peaks in filter output over
some time interval, and then comparing the triggers between each detector to
see if they are consistent with a signal~\cite{babak2013searching,
usman2016pycbc, messick2017analysis}.  The simplest implementations of this
procedure impose that triggers occur within the gravitational wave propagation
time between detectors, but don't otherwise place constraints on the amplitude
or phase measured in each detector.  However, not all gravitational-wave
arrival times are probable or even possible when considering a network of
gravitational wave detectors. Furthermore, arrival times are strongly
correlated with the measured amplitude and phase of the gravitational wave
signals.  Ignoring these effects means that one increases the background
(false-positive rate) of a given gravitational-wave search by considering
unphysical gravitational wave signals.

Work to check consistency of the distribution of measured amplitude, time and
phase of gravitational wave triggers from a network of detectors has been
explored in several contexts.  The early work used only one-dimensional
marginalized distributions for these parameters~\cite{cannon2008bayesian},
which did not accurately describe the joint probability for the network but was
nevertheless demonstrably more effective than simply using coarse windows in
arrival time to define coincidences.  A joint probability distribution of
signal amplitudes for the LIGO and Virgo network was developed for compact
binary searches in ~\cite{2015likelihoodratio} and used during advanced LIGO's
first observing run, however time of arrival and phase parameters were ignored.
Explicitly constructing probabilities for time, phase and amplitude for the
two LIGO detectors and using this information in ranking of candidate events
was explored in the second Advanced LIGO observing
run~\cite{nitz2017detecting}.

In this work, we present a method for accounting for signal consistency in
compact binary searches across a network of gravitational wave detectors that
is computationally efficient and will scale to the eventual 5 detector
gravitational wave network.  This is accomplished through a factorization of
the likelihood ratio (LR) ranking statistic first described in
~\cite{2015likelihoodratio} and used in the analysis of candidates from the
compact binary coalescence search in the first observing run
O1~\cite{messick2017analysis, 2018sachdev, abbott2016binary,
abbott2016gw151226, abbott2016gw150914, abbott2016observation} that generalizes
to multiple gravitational wave detectors, and which takes into account the
correlated trigger amplitudes, times and phases between multiple gravitational
wave observatories.  While the main objective of this work is to document a
novel gravitational wave detection algorithm, it also serves as a reference for
methods used in LIGO and Virgo searches in Advanced LIGO's second observing run
and advanced Virgo's first observing run~\cite{ligo2018gwtc}.

\section{The Likelihood Ratio} \label{sec:llhratio}

In this section we describe the likelihood ratio (LR) ranking statistic as
implemented in the \gstlal inspiral pipeline~\cite{messick2017analysis,
2018sachdev, ligo2018gstlal} with an emphasis on terms relevant to this
particular discussion.  In brief, the \gstlal inspiral pipeline analyzes data
from multiple gravitational wave detectors with time-domain matched
filtering~\cite{cannon2012toward} over a collection of gravitational wave
templates.  First, the peaks in each template filter output are identified for
each detector.  Then, these peaks, which are called triggers, are combined to
look for coincident triggers in the gravitational wave network.  If triggers from
the same template are within the gravitational wave propagation time between
observatories, the collection of triggers is called a coincident
event~\cite{privitera2014improving}.  

Coincident events are common and the vast majority are not gravitational waves.
This is due to the fact that the threshold in each detector for identifying a
trigger is very low leading to potentially billions of triggers per analysis
and tens of millions found in coincidence. Therefore, we have to rank the
collection of coincident events from least to most probable of being a
gravitational wave.  To do this, we explicitly evaluate the LR over the
trigger parameters.  We consider a LR ranking statistic for gravitational-wave
candidates defined as~\footnote{This only includes terms relevant to this
paper.}
\begin{align} \label{eq:lnL}
	\likelihood &= \frac{ \cprob{\Dh, \Ob, \SNR, \CHISQ, \PHI, \T}{\sh} }
			   { \cprob{\Dh, \Ob, \SNR, \CHISQ, \PHI, \T}{\nh} } .
\end{align}
Each vector of parameters is used to denote detector specific information.  To
be concrete we will assume that LIGO Hanford, $H1$, LIGO Livingston, $L1$, and
Virgo, $V1$ are all in observing mode and being analyzed.  However, we note that
the method described can be generalized to add more detectors.  $\Dh$ is a
vector of horizon distances for each observatory $\vec{D_H}=\{D_\mathrm{H1},
D_\mathrm{L1},D_\mathrm{V1}\}$, which accounts for how sensitive the detectors
are at the time of the event.  $\Ob$ is the set of detectors that observed the
event in coincidence.  Since it is possible for a gravitational wave to not
register in one or more detectors above threshold, $\Ob$ will take on one of
six possible values in this case: $\Ob \in \{ \{H1, L1, V1\}$ , $\{H1, L1\}$,
$\{H1, V1\}$, $\{L1, V1\}$, $\{H1\}$, $\{L1\}$, $\{V1\}\}$. $\SNR$ is the
vector individual signal-to-noise ratios (SNRs) in each detector, e.g.,
$\vec{\rho}=\{\rho_\mathrm{H1}, \rho_\mathrm{L1}, \rho_\mathrm{V1}\}$ and
$\CHISQ$ is the vector of $\xi^2$-signal-based-veto values for each detector
(described in~\cite{messick2017analysis}), e.g.,
$\vec{\xi^2}=\{\xi^2_\mathrm{H1}, \xi^2_\mathrm{L1}, \xi^2_\mathrm{V1}\}$.
Likewise, $\T$ and $\PHI$ are the time and phases measured at each
gravitational wave detector~\cite{allen2012findchirp}, $\T=\{t_\mathrm{H1},
t_\mathrm{L1}, t_\mathrm{V1}\}$ and $\PHI=\{\phi_\mathrm{H1}, \phi_\mathrm{L1},
\phi_\mathrm{V1} \}$.

The independence between detectors for noise events implies that the
denominator of \eqref{eq:lnL}, $\cprob{\ldots}{\nh}$, factors to the product of
one-dimensional and two-dimensional distributions and is therefore generally
easy to estimate and evaluate numerically. The numerator, $\cprob{\ldots}{\sh}$
is not easily factorable.  Reference~\cite{2015likelihoodratio} outlines the
factorization of Eq.~\ref{eq:lnL} without the $\vec{\Delta \phi}$ and
$\vec{\Delta t}$ terms, though we point out that it does include the joint
distribution of SNR.  This form was used in the ranking of gravitational-wave
candidates from the first observing run, O1, during which only the H1 and L1
advanced LIGO detectors were
operating~\cite{abbott2016observation,abbott2016gw151226}.  Here we extend the
technique used in~\cite{2015likelihoodratio} and propose the following
factorization of the numerator,
\begin{align}
\cprob{\ldots}{\sh} &= \cprob{\Dh}{\sh} \nonumber \\
&\times \cprob{\Ob}{\Dh, \sh} \nonumber \\
&\times \cprob{\CHISQ}{\SNR, \sh} \nonumber \\
&\times \cprob{\SNR, \PHI, \T}{\Ob, \Dh, \sh},
\end{align}
where we assume that the distribution of the $\CHISQ$-signal-based-veto values
are independent of $\PHI$ and $\T$ but not $\SNR$.

In this work we are concerned with an approximation to the final term:
$\cprob{\SNR, \PHI, \T}{\Ob, \Dh, \sh}$, which accounts for the amplitude,
time, and phase measured in each gravitational wave detector. We begin by
isolating an overall term that scales with the detector network SNR,
$\magnitude{\SNR}$, to the negative fourth power~\cite{schutz2011networks} so
that
\begin{multline}
\label{eq:dtdphi2}
\cprob{\SNR, \PHI, \T}{\Ob, \Dh, \sh} \approx \\
\cprob{\SNR / \magnitude{\SNR}, \PHI, \T}{\Ob, \Dh, \sh} \times \magnitude{\SNR}^{-4}
\end{multline}
This approximation ignores the fact that the accuracy of the measured $\PHI$
and $\T$ depend on the SNR, $\SNR$. In other words, at low SNR the distribution
of time and phase is broad and at high SNR it is narrow. However, since it is
most critical for the detection process to catch signals that are near
threshold, we assume that the uncertainty in $\T$ and $\PHI$ take on values
consistent with a network SNR $\sim 10$.  Thus, our goal is to adequately describe the
$\cprob{\SNR / \magnitude{\SNR}, \PHI, \T}{\Ob, \Dh, \sh}$ when the network SNR
is about 10. 

Next, we note that the absolute time and phase of the signal is arbitrary.
Assuming isotropically distributed gravitational waves, the distribution of
time and phase at a given detector is uniform. However, the relative times and
phases between detectors make up a nontrivial correlated probability density
function. This suggests that we can reduce the dimensionality by computing
parameters relative to a fiducial instrument either by difference or ratio. We
pick the first instrument in alphabetical order.
Furthermore, since $\cprob{\SNR / \magnitude{\SNR}, \PHI, \T}{\Ob, \Dh, \sh}$
is conditional on the detector horizon distances, we can evaluate the effective
distances, $\Deff=\{d_\mathrm{H1}, d_\mathrm{L1}, d_\mathrm{V1} \}$~\cite{allen2012findchirp} instead of $\SNR$.  The advantage to
switching to effective distances, is that the distribution of effective distance
ratios is constant.  Thus we have,
\begin{multline}
\label{eq:dtdphi2}
\cprob{\SNR, \PHI, \T}{\Ob, \Dh, \sh} \overset{\propto}{\sim} \\
\cprob{\DDeff, \DPHI, \DT}{\Ob, \sh} \times \magnitude{\SNR}^{-4},
\end{multline}
where concretely for the $H1$, $L1$, $V1$ network
\begin{align}
\DDeff &\equiv \{\ln(d_{L1} / d_{H1}), \ln(d_{V1} / d_{H1})\} \nonumber, \\
\DT &\equiv \{t_{L1} - t_{H1}, t_{V1} - t_{H1}\} \nonumber, \\
\DPHI &\equiv \{\phi_{L1} - \phi_{H1}, \phi_{V1} - \phi_{H1}\} \nonumber .
\end{align}
In interest of simplifying the subsequent notation, we define a single vector of
parameters that contains $\DDeff$, $\DT$ and $\DPHI$,
\begin{align}
\Ext &\equiv \{\DDeff, \DT, \DPHI \}.
\end{align}

In order to construct the distribution of these parameters for a signal, we
assert that gravitational waves have a uniform distribution in Earth-based
coordinates: right ascension $\alpha$, declination $\cos(\delta)$, inclination
angle $\cos(\iota)$, and polarization angle $\psi$.  We lay down a uniform,
densely sampled grid in $\{\alpha, \cos(\delta), \cos(\iota), \psi\}$ and
further assert that any signal should ``exactly" land on one of the grid
points. We transform that regularly-spaced grid into a grid of irregularly
spaced points in $\Ext$, which we denote as $\Extmod$ for the $i^{th}$ model
vector.  We consider that the only mechanism to push a signal away from one of
these exact $i^{th}$ grid points is Gaussian noise.  Furthermore we assume the
probability distribution is of the form:
\begin{multline} \label{eq:dtdphi4}
\cprob{\Ext}{\Ob, \sh, \Extmod} = \\
\frac{1}{\sqrt{(2\pi)^k |\Cov_{\Ext}|}} \exp{ \left[ -\frac{1}{2} \Dext_i \Cov^{-1}_{\Ext} \Dext_i^T \right] },
\end{multline}
where $\Dext_i \equiv \Ext - \Extmod$ and $ \Cov_{\Ext} $ is a covariance matrix of $ \Ext $. We further assume that the measurement of the time, phase, effective distance for an individual detector is independent of that for the other detectors. Therefore, every element of $ \Ext $ can be expressed as
\begin{align}
\Sigma_{ij}={\sigma^2}_{\lambda_{i}\lambda_{j}} = {\sigma^2}_{\theta_i \theta_j}^{(\mathrm{ifo}1)} + {\sigma^2}_{\theta_i \theta_j}^{(\mathrm{ifo}2)},
\end{align}
where $ \vec{\theta} $ is a vector of the three extrinsic parameters of interest, namely $ \vec{\theta}\equiv\{\ln D_{\textrm{eff}}, t, \phi\}$, and the superscripts (ifo1/2) indicate an individual detector in a pair, at which the parameters are measured.
The covariance matrix on $ \vec{\theta}$ can be approximated by the inverse of a Fisher information matrix. Hence, one can obtain the covariance matrix as follows:
\begin{align}
\label{Cov_mat}
\Cov_{\vec{\theta}} &\equiv \begin{bmatrix}
{\sigma^2}_{tt} & {\sigma^2}_{t\phi}& {\sigma}^2_{t  \ln D_{\textrm{eff}}}\\
{\sigma^2}_{\phi t}& {\sigma^2}_{\phi  \phi}& {\sigma^2}_{\phi  \ln D_{\textrm{eff}}}  \\
{\sigma^2}_{\ln D_{\textrm{eff}} \, t} & {\sigma^2}_{\ln D_{\textrm{eff}} \, \phi} & {\sigma^2}_{\ln  D_{\textrm{eff}}  \ln D_{\textrm{eff}}}
\end{bmatrix}\\
&=\begin{bmatrix}
\frac{1}{\big(2\pi\, \rho \, \sigma_f \big)^2} & \frac{\bar{f}}{2\pi\,\rho^2\,\sigma_f^2} & 0\\
\frac{\bar{f}}{2\pi\,\rho^2\,\sigma_f^2} & \frac{\bar{f^2}}{\big(\rho \, \sigma_f \big)^2} & 0  \\
0 & 0 & \frac{1}{\rho^2} 
\end{bmatrix}.
\end{align}
$\sigma_f$ is the effective bandwidth of the signal, defined as \cite{fairhurst2009}
\begin{align}
\sigma_{f}^2 \equiv \bar{f^2} - \left(\bar{f^1}\right)^2,
\end{align}
using the frequency moments of the signal
\begin{align}\label{eq:fmoment}
\bar{f^n} \equiv 4 \int_{0}^{\infty} df \frac{|\tilde{h}(f)|^2}{S(f)} f^n.
\end{align}

From Eq. \ref{Cov_mat}, it is clear that the covariance matrix elements depend on the observed SNR at each detector. Since the goal of the present work is to improve efficiency of detecting near threshold triggers, we have set the characteristic SNR for Hanford, Livingston and Virgo detector as 5, 7 and 2.25 respectively.

The inverse of the covariance matrix $\Cov_{\Ext}$ further can be decomposed into following two matrices using Cholesky decomposition \cite{golub2012}.
\begin{equation}
 \big(\Cov_{\Ext}\big)^{-1} = \bf{\mathcal{C}}\,\bf{\mathcal{C}}^T,
\end{equation}
where $\bf{\mathcal{C}}$ is a lower triangular matrix.
The matrix $\bf{\mathcal{C}}$ is then used to obtain a re-scaled set of orthogonal coordinates $\vec{x}$ such that $\DX_i \equiv \Big(\Ext - \Extmod\Big)\,\mathcal{C} =  \Delta \vec{\lambda}_{i} \, \mathcal{C}$. Hence, the probability distribution described in the Eq.\eqref{eq:dtdphi4} can be rewritten in terms of the new coordinates as
\begin{equation}
P\Big(\vec{\lambda}| \vec{O}, s, \vec{\lambda}_{\textrm{mi}} \Big) \propto \textrm{exp} \Big[ - \frac{1}{2} \vec{\Delta x_i}^2 \Big]
\end{equation}

\begin{figure}[h!]
\centering
\includegraphics[width=0.45\textwidth]{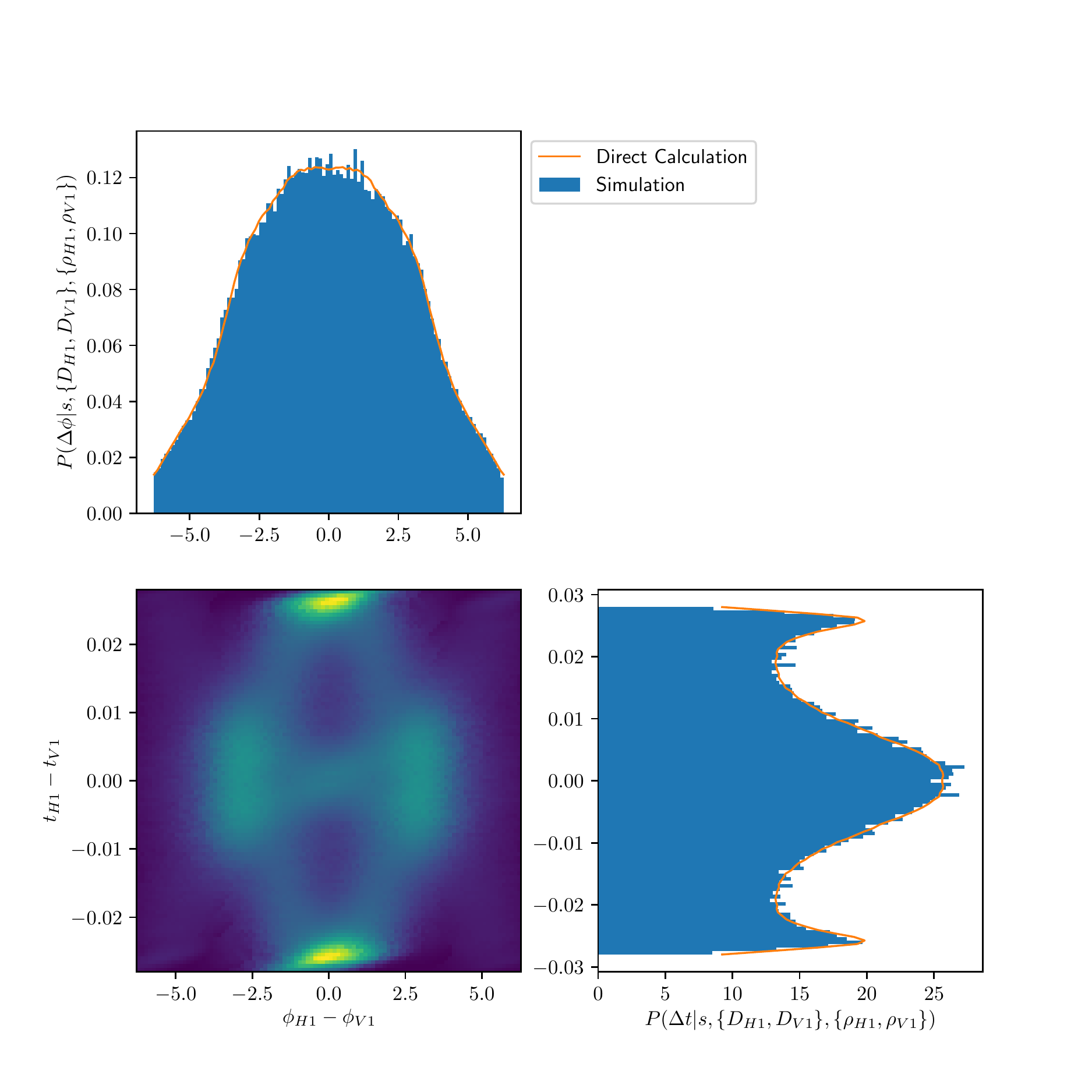}
\caption{Example probability density function for the difference in
gravitational wave arrival time and phase between LIGO Hanford and Virgo. Here
we set the horizon distances of LIGO Hanford and Virgo to be $ 110~\mathrm{Mpc} $ and $ 45~\mathrm{Mpc} $ respectively to be consistent with realistic PSDs of the two detectors, which are used for \eqref{eq:fmoment}, and show a
slice of the probability density function where the ratios of the effective distance are close to 1.
The full two dimensional distribution (lower left) was computed using the
method described in section \ref{sec:llhratio}. The orange traces use the data from the two dimensional distribution marginalized over
phase and time respectively. For comparison, we computed the marginal
distributions via a direct Monte Carlo method to obtain the top and bottom right bar
plots. The agreement is excellent, however, we note that the same error assumptions were
used in the Monte Carlo method and that changing the assumptions about the
covariance matrix will lead to poorer agreement.}
\label{fig:examplePDF}
\end{figure}

\begin{figure}[h!]
	\centering
	\includegraphics[width=0.45\textwidth]{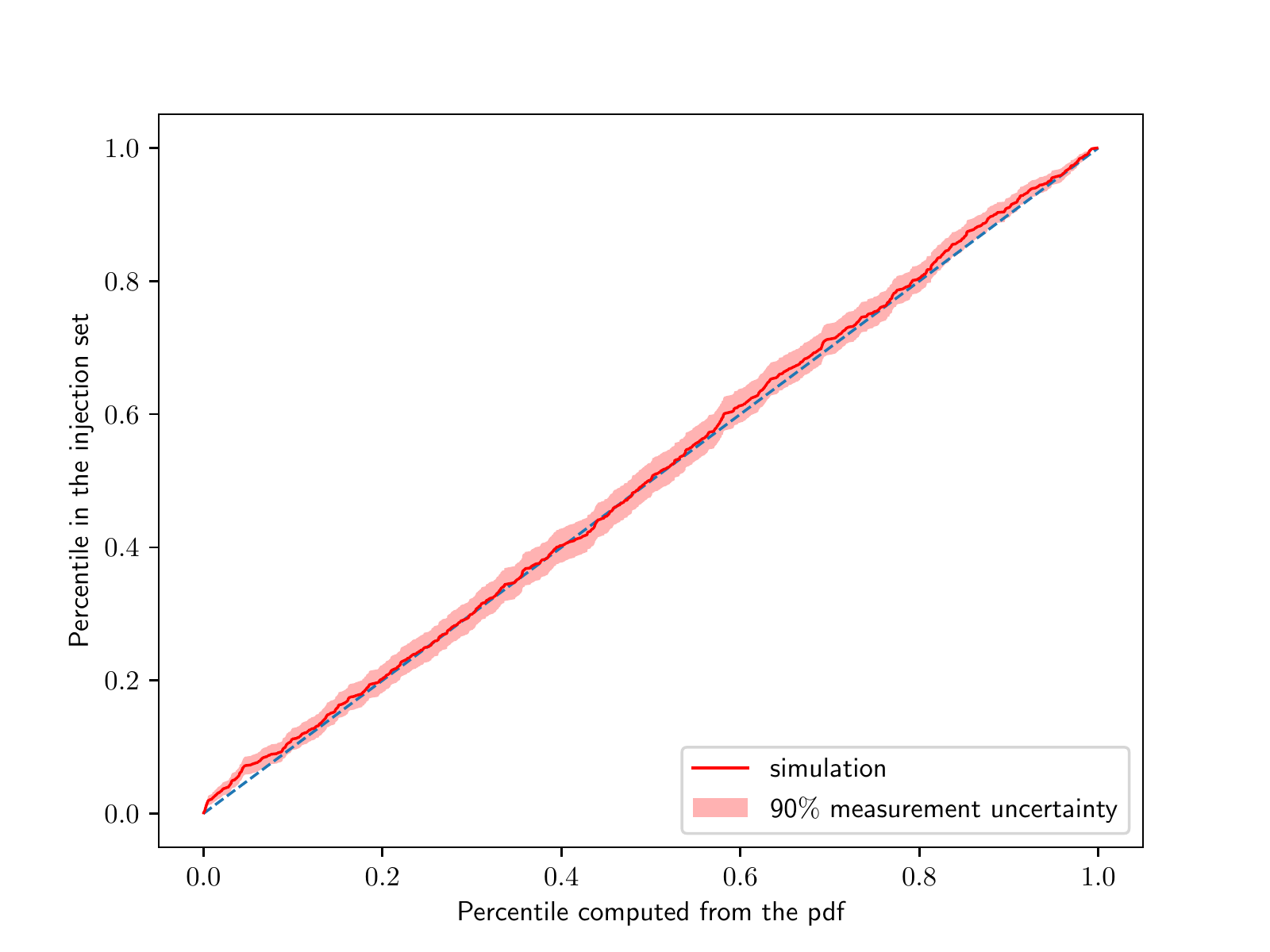}
	\caption{P-p plot created by a subset of the Monte Carlo samples shown in Fig.~\ref{fig:examplePDF}. The uncertainty on the measured percentiles due to a finite number of samples is shown as shaded region. We find that the diagonal line sits in the error region in the entire percentile random which indicates an agreement between the samples and the two dimensional distribution.} 
	\label{fig:pp_plot}
\end{figure}

Then we assert that
\begin{align} \label{eq:dtdphi6}
\cprob{\Ext}{\Ob, \sh} &\propto \sum_i \cprob{\Ext}{\Ob, \sh, \Extmod} \prob{\Extmod} \nonumber \\
	&\propto \sum_i \cprob{\Ext}{\Ob, \sh, \Extmod} \nonumber \\
	&= \sum_i \exp{ \left[ -\frac{1}{2} \DX_i^2 \right] },
\end{align}
where the second line holds by construction since $\prob{\Extmod}$ does
not depend on $i$ as they were chosen uniform in prior signal probability.
Computing probability \eqref{eq:dtdphi6} in real-time for each gravitational
wave candidate is not feasible since the grid might have millions of points.
Therefore, we make another simplifying assumption: we assume that the noise
only adds a contribution which is orthogonal to the hyper-surface defined by the
signal.  In other words, we assume that noise cannot push a signal from one
grid point towards another along the signal hyper-surface which implies,
\begin{equation}
\DX_i^2 \approx \DX_0^2 + g_{0i}^2,
\end{equation}
where $\DX_0$ refers to the distance between the candidate parameter and the
\textit{nearest-neighbor} grid point, and $g_{0i}$ is the distance between the
$i$th grid point and the nearest neighbor grid point. In this case we can
simplify the marginalization step with a precomputation since
\begin{align} \label{eq:dtdphi7}
\cprob{\Ext}{\Ob, \sh} &\approx \exp{ \left[ -\frac{1}{2} \DX_0^2 \right]} \nonumber \\
&\times \sum_i \exp{ \left[ -\frac{1}{2} g_{0i}^2 \right]}.
\end{align}
The entire sum over $i$ can be precomputed and stored.  In order for this
endeavor to be successful, we still need a fast way to find the nearest
neighbor. We use SciPy KDTree to accomplish this~\cite{jones2014scipy}.
Fig.~\ref{fig:examplePDF} shows an example probability density function for
time delays and phase differences between LIGO Hanford and Virgo computed from
the above method as well as a comparison with $1\times10^5$ samples produced by a direct Monte Carlo. Fig.~\ref{fig:pp_plot} is a probability-probability (p-p) plot created by a subset of the samples shown in Fig.~\ref{fig:examplePDF}. We find that the drawn samples are in a good agreement with the two dimensional distribution over phase and time. We note that each evaluation of \eqref{eq:dtdphi7} required less than a millisecond to
compute on a modern CPU. Thus, this method is suitable for evaluating hundreds
of candidates per second per CPU core in real-time.

\section{Results} \label{sec:results}

To assess the effect of the procedure described in \ref{sec:llhratio}, we
conducted a simulation of synthetic signals and non-stationary noise in order to
produce a Receiver Operating Characteristics (ROC) curve. We constructed
synthetic signal and noise models. For the signal model, we assumed sources
uniform in the volume of space and isotropically oriented.  We constructed
$1\times10^7$ such signals and calculated their corresponding SNRs, arrival times, and
phases in the LIGO Hanford and LIGO Livingston detectors using the LAL
Simulation package~\cite{ligo2018lalsuite}.  For noise, we constructed $1\times10^7$
simulated glitches with SNRs that were independent in each of the LIGO Hanford
and LIGO Livingston detectors given by an exponential distribution 
\begin{equation}
\prob{\rho_{H1}, \rho_{L1}} = \frac{1}{6}\exp(-\frac{\rho_{H1}}{6}) \, \frac{1}{6}\exp(-\frac{\rho_{L1}}{6}),
\end{equation}
where we placed an additional constraint that we only kept samples which had
both $\rho_{H1} \geq 4$ and $\rho_{L1} \geq 4$.  We chose arrival time differences
that were uniform within the GW travel time between the two detectors and phase
differences that were uniform between 0 and $2\pi$.

\begin{figure}[h!]
	\centering
\includegraphics[width=.45\textwidth]{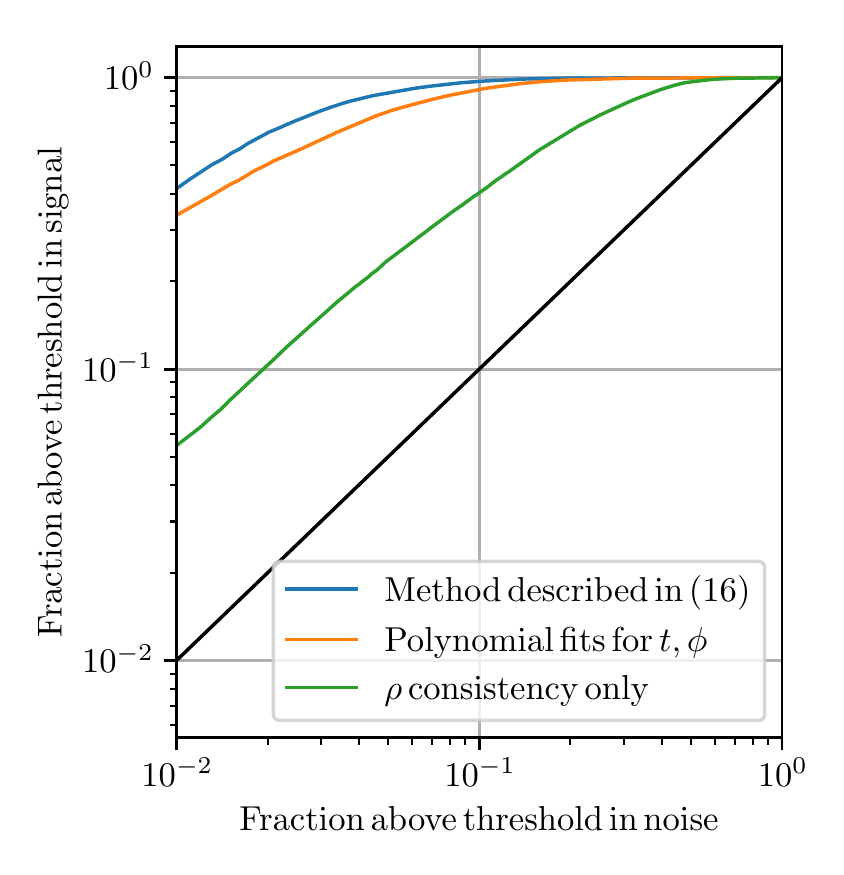}
\caption{ROC curve showing performance of signal consistency check from the
simulation described in \ref{sec:results}. The ROC curve plots the fraction of
simulated events found above a given threshold of LR vs the fraction of noise
found above that threshold.  The black diagonal line would indicate that the
method is not helpful at discriminating signals from noise.  It is desirable to
have the performance of any method be above the black diagonal line. The green
curve shows the likelihood ratio including only the amplitude consistency terms
($\SNR$) used in the first advanced LIGO observing
run~\cite{2015likelihoodratio}. The orange curve shows the inclusion of $\DT$
and $\DPHI$ terms used in the beginning of advanced LIGO's second observing
run~\cite{2018sachdev} and the blue curve shows method described in
section \ref{sec:llhratio}.
\label{fig:roc}
}
\end{figure}

Figure~\ref{fig:roc} shows the ROC curves for three cases: the SNR-only terms
in the LR, which was used in initial advanced LIGO
searches~\cite{abbott2016observation, abbott2016gw151226, abbott2016binary}
(green); a previous implementation of time and phase consistency used for
subsequent gravitational wave searches~\cite{2018sachdev,
scientific2017gw170104, abbott2017gw170608, abbott2017gw170814,
abbott2017gw170817} terms (orange); and finally the current implementation described in
section \ref{sec:llhratio} (blue). Here we see that even in the two-detector situation, the current implementation achieves an improvement for the false alarm probability above $ 1\times10^{-3} $.

\section{Conclusion} \label{sec:conclusion}

We have demonstrated a computationally efficient likelihood ratio statistic for
use in gravitational wave searches for compact binary systems with LIGO and
Virgo.  The work here specifically presents a way to assess the probability
that a given set of measurements independently made at each observatory in a
worldwide network of observatories is consistent with a signal.  This is useful
for real-time identification of candidates where the latency of the computation
is critical and the overall computational scale cannot be too large.  Our
method allows for computing the likelihood ratio for hundreds of candidates per
second on a single modern CPU core.  This method was used to produce the final
results for advanced LIGO's second observing run and Advanced Virgo's first
observing run with the \gstlal pipeline.

\section{Acknowledgements} \label{sec:ack}

This work was supported by the National Science Foundation through PHY-1454389,
OAC-1841480, ACI-1642391, PHY-1700765, and PHY-1607585.  Funding for this
project was provided by the Charles E.  Kaufman Foundation of The Pittsburgh
Foundation.  We thank the LIGO Scientific Collaboration for input on this work.
Specifically, CH would like to thank Patrick Brady for several illuminating
discussions.  This research was supported in part by Perimeter Institute for
Theoretical Physics. Research at Perimeter Institute is supported by the
Government of Canada through the Department of Innovation, Science, and
Economic Development, and by the Province of Ontario through the Ministry of
Research and Innovation.  Computations for this research were performed on the
Pennsylvania State University’s Institute for CyberScience Advanced
CyberInfrastructure (ICS-ACI).  We are grateful for computational resources
provided by the Leonard E Parker Center for Gravitation, Cosmology and
Astrophysics at the University of Wisconsin-Milwaukee and supported by National
Science Foundation Grants PHY-1626190 and PHY-1700765.  The authors are
grateful for computational resources provided by the LIGO Laboratory and
supported by National Science Foundation Grants PHY-0757058 and PHY-0823459.
This paper has LIGO document number: P1800362.

\appendix \label{sec:appendix}

\bibliography{references}

\begin{thebibliography}{42}%
\makeatletter
\providecommand \@ifxundefined [1]{%
 \@ifx{#1\undefined}
}%
\providecommand \@ifnum [1]{%
 \ifnum #1\expandafter \@firstoftwo
 \else \expandafter \@secondoftwo
 \fi
}%
\providecommand \@ifx [1]{%
 \ifx #1\expandafter \@firstoftwo
 \else \expandafter \@secondoftwo
 \fi
}%
\providecommand \natexlab [1]{#1}%
\providecommand \enquote  [1]{``#1''}%
\providecommand \bibnamefont  [1]{#1}%
\providecommand \bibfnamefont [1]{#1}%
\providecommand \citenamefont [1]{#1}%
\providecommand \href@noop [0]{\@secondoftwo}%
\providecommand \href [0]{\begingroup \@sanitize@url \@href}%
\providecommand \@href[1]{\@@startlink{#1}\@@href}%
\providecommand \@@href[1]{\endgroup#1\@@endlink}%
\providecommand \@sanitize@url [0]{\catcode `\\12\catcode `\$12\catcode
  `\&12\catcode `\#12\catcode `\^12\catcode `\_12\catcode `\%12\relax}%
\providecommand \@@startlink[1]{}%
\providecommand \@@endlink[0]{}%
\providecommand \url  [0]{\begingroup\@sanitize@url \@url }%
\providecommand \@url [1]{\endgroup\@href {#1}{\urlprefix }}%
\providecommand \urlprefix  [0]{URL }%
\providecommand \Eprint [0]{\href }%
\providecommand \doibase [0]{http://dx.doi.org/}%
\providecommand \selectlanguage [0]{\@gobble}%
\providecommand \bibinfo  [0]{\@secondoftwo}%
\providecommand \bibfield  [0]{\@secondoftwo}%
\providecommand \translation [1]{[#1]}%
\providecommand \BibitemOpen [0]{}%
\providecommand \bibitemStop [0]{}%
\providecommand \bibitemNoStop [0]{.\EOS\space}%
\providecommand \EOS [0]{\spacefactor3000\relax}%
\providecommand \BibitemShut  [1]{\csname bibitem#1\endcsname}%
\let\auto@bib@innerbib\@empty
\bibitem [{\citenamefont {Harry}\ \emph {et~al.}(2010)\citenamefont {Harry},
  \citenamefont {Collaboration} \emph {et~al.}}]{harry2010advanced}%
  \BibitemOpen
  \bibfield  {author} {\bibinfo {author} {\bibfnamefont {G.~M.}\ \bibnamefont
  {Harry}}, \bibinfo {author} {\bibfnamefont {L.~S.}\ \bibnamefont
  {Collaboration}},  \emph {et~al.},\ }\href@noop {} {\bibfield  {journal}
  {\bibinfo  {journal} {Classical and Quantum Gravity}\ }\textbf {\bibinfo
  {volume} {27}},\ \bibinfo {pages} {084006} (\bibinfo {year}
  {2010})}\BibitemShut {NoStop}%
\bibitem [{\citenamefont {Aasi}\ \emph {et~al.}(2015)\citenamefont {Aasi},
  \citenamefont {Abbott}, \citenamefont {Abbott}, \citenamefont {Abbott},
  \citenamefont {Abernathy}, \citenamefont {Ackley}, \citenamefont {Adams},
  \citenamefont {Adams}, \citenamefont {Addesso}, \citenamefont {Adhikari}
  \emph {et~al.}}]{aasi2015advanced}%
  \BibitemOpen
  \bibfield  {author} {\bibinfo {author} {\bibfnamefont {J.}~\bibnamefont
  {Aasi}}, \bibinfo {author} {\bibfnamefont {B.}~\bibnamefont {Abbott}},
  \bibinfo {author} {\bibfnamefont {R.}~\bibnamefont {Abbott}}, \bibinfo
  {author} {\bibfnamefont {T.}~\bibnamefont {Abbott}}, \bibinfo {author}
  {\bibfnamefont {M.}~\bibnamefont {Abernathy}}, \bibinfo {author}
  {\bibfnamefont {K.}~\bibnamefont {Ackley}}, \bibinfo {author} {\bibfnamefont
  {C.}~\bibnamefont {Adams}}, \bibinfo {author} {\bibfnamefont
  {T.}~\bibnamefont {Adams}}, \bibinfo {author} {\bibfnamefont
  {P.}~\bibnamefont {Addesso}}, \bibinfo {author} {\bibfnamefont
  {R.}~\bibnamefont {Adhikari}},  \emph {et~al.},\ }\href@noop {} {\bibfield
  {journal} {\bibinfo  {journal} {Classical and quantum gravity}\ }\textbf
  {\bibinfo {volume} {32}},\ \bibinfo {pages} {074001} (\bibinfo {year}
  {2015})}\BibitemShut {NoStop}%
\bibitem [{\citenamefont {Acernese}\ \emph {et~al.}(2014)\citenamefont
  {Acernese}, \citenamefont {Agathos}, \citenamefont {Agatsuma}, \citenamefont
  {Aisa}, \citenamefont {Allemandou}, \citenamefont {Allocca}, \citenamefont
  {Amarni}, \citenamefont {Astone}, \citenamefont {Balestri}, \citenamefont
  {Ballardin} \emph {et~al.}}]{acernese2014advanced}%
  \BibitemOpen
  \bibfield  {author} {\bibinfo {author} {\bibfnamefont {F.}~\bibnamefont
  {Acernese}}, \bibinfo {author} {\bibfnamefont {M.}~\bibnamefont {Agathos}},
  \bibinfo {author} {\bibfnamefont {K.}~\bibnamefont {Agatsuma}}, \bibinfo
  {author} {\bibfnamefont {D.}~\bibnamefont {Aisa}}, \bibinfo {author}
  {\bibfnamefont {N.}~\bibnamefont {Allemandou}}, \bibinfo {author}
  {\bibfnamefont {A.}~\bibnamefont {Allocca}}, \bibinfo {author} {\bibfnamefont
  {J.}~\bibnamefont {Amarni}}, \bibinfo {author} {\bibfnamefont
  {P.}~\bibnamefont {Astone}}, \bibinfo {author} {\bibfnamefont
  {G.}~\bibnamefont {Balestri}}, \bibinfo {author} {\bibfnamefont
  {G.}~\bibnamefont {Ballardin}},  \emph {et~al.},\ }\href@noop {} {\bibfield
  {journal} {\bibinfo  {journal} {Classical and Quantum Gravity}\ }\textbf
  {\bibinfo {volume} {32}},\ \bibinfo {pages} {024001} (\bibinfo {year}
  {2014})}\BibitemShut {NoStop}%
\bibitem [{\citenamefont {Abbott}\ \emph
  {et~al.}(2016{\natexlab{a}})\citenamefont {Abbott}, \citenamefont {Abbott},
  \citenamefont {Abbott}, \citenamefont {Abernathy}, \citenamefont {Acernese},
  \citenamefont {Ackley}, \citenamefont {Adams}, \citenamefont {Adams},
  \citenamefont {Addesso}, \citenamefont {Adhikari} \emph
  {et~al.}}]{abbott2016observation}%
  \BibitemOpen
  \bibfield  {author} {\bibinfo {author} {\bibfnamefont {B.~P.}\ \bibnamefont
  {Abbott}}, \bibinfo {author} {\bibfnamefont {R.}~\bibnamefont {Abbott}},
  \bibinfo {author} {\bibfnamefont {T.}~\bibnamefont {Abbott}}, \bibinfo
  {author} {\bibfnamefont {M.}~\bibnamefont {Abernathy}}, \bibinfo {author}
  {\bibfnamefont {F.}~\bibnamefont {Acernese}}, \bibinfo {author}
  {\bibfnamefont {K.}~\bibnamefont {Ackley}}, \bibinfo {author} {\bibfnamefont
  {C.}~\bibnamefont {Adams}}, \bibinfo {author} {\bibfnamefont
  {T.}~\bibnamefont {Adams}}, \bibinfo {author} {\bibfnamefont
  {P.}~\bibnamefont {Addesso}}, \bibinfo {author} {\bibfnamefont
  {R.}~\bibnamefont {Adhikari}},  \emph {et~al.},\ }\href@noop {} {\bibfield
  {journal} {\bibinfo  {journal} {Physical review letters}\ }\textbf {\bibinfo
  {volume} {116}},\ \bibinfo {pages} {061102} (\bibinfo {year}
  {2016}{\natexlab{a}})}\BibitemShut {NoStop}%
\bibitem [{\citenamefont {Abbott}\ \emph
  {et~al.}(2018{\natexlab{a}})\citenamefont {Abbott} \emph
  {et~al.}}]{ligo2018gwtc}%
  \BibitemOpen
  \bibfield  {author} {\bibinfo {author} {\bibfnamefont {B.~P.}\ \bibnamefont
  {Abbott}} \emph {et~al.},\ }\href@noop {} {\bibfield  {journal} {\bibinfo
  {journal} {arXiv preprint arXiv:1811.12907}\ } (\bibinfo {year}
  {2018}{\natexlab{a}})}\BibitemShut {NoStop}%
\bibitem [{\citenamefont {Abbott}\ \emph
  {et~al.}(2018{\natexlab{b}})\citenamefont {Abbott}, \citenamefont {Abbott},
  \citenamefont {Abbott} \emph {et~al.}}]{aasi2016prospects}%
  \BibitemOpen
  \bibfield  {author} {\bibinfo {author} {\bibfnamefont {B.}~\bibnamefont
  {Abbott}}, \bibinfo {author} {\bibfnamefont {R.}~\bibnamefont {Abbott}},
  \bibinfo {author} {\bibfnamefont {T.}~\bibnamefont {Abbott}},  \emph
  {et~al.},\ }\href@noop {} {\bibfield  {journal} {\bibinfo  {journal} {Living
  Reviews in Relativity}\ }\textbf {\bibinfo {volume} {21}} (\bibinfo {year}
  {2018}{\natexlab{b}})}\BibitemShut {NoStop}%
\bibitem [{\citenamefont {Somiya}(2012)}]{somiya2012detector}%
  \BibitemOpen
  \bibfield  {author} {\bibinfo {author} {\bibfnamefont {K.}~\bibnamefont
  {Somiya}},\ }\href@noop {} {\bibfield  {journal} {\bibinfo  {journal}
  {Classical and Quantum Gravity}\ }\textbf {\bibinfo {volume} {29}},\ \bibinfo
  {pages} {124007} (\bibinfo {year} {2012})}\BibitemShut {NoStop}%
\bibitem [{\citenamefont {Aso}\ \emph {et~al.}(2013)\citenamefont {Aso},
  \citenamefont {Michimura}, \citenamefont {Somiya}, \citenamefont {Ando},
  \citenamefont {Miyakawa}, \citenamefont {Sekiguchi}, \citenamefont {Tatsumi},
  \citenamefont {Yamamoto}, \citenamefont {Collaboration} \emph
  {et~al.}}]{aso2013interferometer}%
  \BibitemOpen
  \bibfield  {author} {\bibinfo {author} {\bibfnamefont {Y.}~\bibnamefont
  {Aso}}, \bibinfo {author} {\bibfnamefont {Y.}~\bibnamefont {Michimura}},
  \bibinfo {author} {\bibfnamefont {K.}~\bibnamefont {Somiya}}, \bibinfo
  {author} {\bibfnamefont {M.}~\bibnamefont {Ando}}, \bibinfo {author}
  {\bibfnamefont {O.}~\bibnamefont {Miyakawa}}, \bibinfo {author}
  {\bibfnamefont {T.}~\bibnamefont {Sekiguchi}}, \bibinfo {author}
  {\bibfnamefont {D.}~\bibnamefont {Tatsumi}}, \bibinfo {author} {\bibfnamefont
  {H.}~\bibnamefont {Yamamoto}}, \bibinfo {author} {\bibfnamefont
  {K.}~\bibnamefont {Collaboration}},  \emph {et~al.},\ }\href@noop {}
  {\bibfield  {journal} {\bibinfo  {journal} {Physical Review D}\ }\textbf
  {\bibinfo {volume} {88}},\ \bibinfo {pages} {043007} (\bibinfo {year}
  {2013})}\BibitemShut {NoStop}%
\bibitem [{\citenamefont {Unnikrishnan}(2013)}]{unnikrishnan2013indigo}%
  \BibitemOpen
  \bibfield  {author} {\bibinfo {author} {\bibfnamefont {C.}~\bibnamefont
  {Unnikrishnan}},\ }\href@noop {} {\bibfield  {journal} {\bibinfo  {journal}
  {International Journal of Modern Physics D}\ }\textbf {\bibinfo {volume}
  {22}},\ \bibinfo {pages} {1341010} (\bibinfo {year} {2013})}\BibitemShut
  {NoStop}%
\bibitem [{\citenamefont {Abbott}\ \emph {et~al.}(2009)\citenamefont {Abbott},
  \citenamefont {Abbott}, \citenamefont {Adhikari}, \citenamefont {Ajith},
  \citenamefont {Allen}, \citenamefont {Allen}, \citenamefont {Amin},
  \citenamefont {Anderson}, \citenamefont {Anderson}, \citenamefont {Arain}
  \emph {et~al.}}]{abbott2009ligo}%
  \BibitemOpen
  \bibfield  {author} {\bibinfo {author} {\bibfnamefont {B.}~\bibnamefont
  {Abbott}}, \bibinfo {author} {\bibfnamefont {R.}~\bibnamefont {Abbott}},
  \bibinfo {author} {\bibfnamefont {R.}~\bibnamefont {Adhikari}}, \bibinfo
  {author} {\bibfnamefont {P.}~\bibnamefont {Ajith}}, \bibinfo {author}
  {\bibfnamefont {B.}~\bibnamefont {Allen}}, \bibinfo {author} {\bibfnamefont
  {G.}~\bibnamefont {Allen}}, \bibinfo {author} {\bibfnamefont
  {R.}~\bibnamefont {Amin}}, \bibinfo {author} {\bibfnamefont {S.}~\bibnamefont
  {Anderson}}, \bibinfo {author} {\bibfnamefont {W.}~\bibnamefont {Anderson}},
  \bibinfo {author} {\bibfnamefont {M.}~\bibnamefont {Arain}},  \emph
  {et~al.},\ }\href@noop {} {\bibfield  {journal} {\bibinfo  {journal} {Reports
  on Progress in Physics}\ }\textbf {\bibinfo {volume} {72}},\ \bibinfo {pages}
  {076901} (\bibinfo {year} {2009})}\BibitemShut {NoStop}%
\bibitem [{\citenamefont {Anderson}\ \emph {et~al.}()\citenamefont {Anderson},
  \citenamefont {Whelan}, \citenamefont {Brady}, \citenamefont {Creighton},
  \citenamefont {Chin},\ and\ \citenamefont {Riles}}]{andersonbeam}%
  \BibitemOpen
  \bibfield  {author} {\bibinfo {author} {\bibfnamefont {W.~G.}\ \bibnamefont
  {Anderson}}, \bibinfo {author} {\bibfnamefont {J.~T.}\ \bibnamefont
  {Whelan}}, \bibinfo {author} {\bibfnamefont {P.~R.}\ \bibnamefont {Brady}},
  \bibinfo {author} {\bibfnamefont {J.~D.}\ \bibnamefont {Creighton}}, \bibinfo
  {author} {\bibfnamefont {D.}~\bibnamefont {Chin}}, \ and\ \bibinfo {author}
  {\bibfnamefont {K.}~\bibnamefont {Riles}},\ }\href
  {https://dcc.ligo.org/public/0012/T010110/001/T010110.pdf} {\ }\BibitemShut
  {NoStop}%
\bibitem [{\citenamefont {Veitch}\ \emph {et~al.}(2015)\citenamefont {Veitch},
  \citenamefont {Raymond}, \citenamefont {Farr}, \citenamefont {Farr},
  \citenamefont {Graff}, \citenamefont {Vitale}, \citenamefont {Aylott},
  \citenamefont {Blackburn}, \citenamefont {Christensen}, \citenamefont
  {Coughlin} \emph {et~al.}}]{veitch2015parameter}%
  \BibitemOpen
  \bibfield  {author} {\bibinfo {author} {\bibfnamefont {J.}~\bibnamefont
  {Veitch}}, \bibinfo {author} {\bibfnamefont {V.}~\bibnamefont {Raymond}},
  \bibinfo {author} {\bibfnamefont {B.}~\bibnamefont {Farr}}, \bibinfo {author}
  {\bibfnamefont {W.}~\bibnamefont {Farr}}, \bibinfo {author} {\bibfnamefont
  {P.}~\bibnamefont {Graff}}, \bibinfo {author} {\bibfnamefont
  {S.}~\bibnamefont {Vitale}}, \bibinfo {author} {\bibfnamefont
  {B.}~\bibnamefont {Aylott}}, \bibinfo {author} {\bibfnamefont
  {K.}~\bibnamefont {Blackburn}}, \bibinfo {author} {\bibfnamefont
  {N.}~\bibnamefont {Christensen}}, \bibinfo {author} {\bibfnamefont
  {M.}~\bibnamefont {Coughlin}},  \emph {et~al.},\ }\href@noop {} {\bibfield
  {journal} {\bibinfo  {journal} {Physical Review D}\ }\textbf {\bibinfo
  {volume} {91}},\ \bibinfo {pages} {042003} (\bibinfo {year}
  {2015})}\BibitemShut {NoStop}%
\bibitem [{\citenamefont {Singer}\ and\ \citenamefont
  {Price}(2016)}]{singer2016rapid}%
  \BibitemOpen
  \bibfield  {author} {\bibinfo {author} {\bibfnamefont {L.~P.}\ \bibnamefont
  {Singer}}\ and\ \bibinfo {author} {\bibfnamefont {L.~R.}\ \bibnamefont
  {Price}},\ }\href@noop {} {\bibfield  {journal} {\bibinfo  {journal}
  {Physical Review D}\ }\textbf {\bibinfo {volume} {93}},\ \bibinfo {pages}
  {024013} (\bibinfo {year} {2016})}\BibitemShut {NoStop}%
\bibitem [{\citenamefont {Pai}\ \emph {et~al.}(2002)\citenamefont {Pai},
  \citenamefont {Bose},\ and\ \citenamefont
  {Dhurandhar}}]{pai2002computational}%
  \BibitemOpen
  \bibfield  {author} {\bibinfo {author} {\bibfnamefont {A.}~\bibnamefont
  {Pai}}, \bibinfo {author} {\bibfnamefont {S.}~\bibnamefont {Bose}}, \ and\
  \bibinfo {author} {\bibfnamefont {S.}~\bibnamefont {Dhurandhar}},\
  }\href@noop {} {\bibfield  {journal} {\bibinfo  {journal} {Classical and
  Quantum Gravity}\ }\textbf {\bibinfo {volume} {19}},\ \bibinfo {pages} {1477}
  (\bibinfo {year} {2002})}\BibitemShut {NoStop}%
\bibitem [{\citenamefont {Macleod}\ \emph {et~al.}(2016)\citenamefont
  {Macleod}, \citenamefont {Harry},\ and\ \citenamefont
  {Fairhurst}}]{macleod2016fully}%
  \BibitemOpen
  \bibfield  {author} {\bibinfo {author} {\bibfnamefont {D.}~\bibnamefont
  {Macleod}}, \bibinfo {author} {\bibfnamefont {I.}~\bibnamefont {Harry}}, \
  and\ \bibinfo {author} {\bibfnamefont {S.}~\bibnamefont {Fairhurst}},\
  }\href@noop {} {\bibfield  {journal} {\bibinfo  {journal} {Physical Review
  D}\ }\textbf {\bibinfo {volume} {93}},\ \bibinfo {pages} {064004} (\bibinfo
  {year} {2016})}\BibitemShut {NoStop}%
\bibitem [{\citenamefont {Klimenko}\ \emph {et~al.}(2008)\citenamefont
  {Klimenko}, \citenamefont {Yakushin}, \citenamefont {Mercer},\ and\
  \citenamefont {Mitselmakher}}]{klimenko2008coherent}%
  \BibitemOpen
  \bibfield  {author} {\bibinfo {author} {\bibfnamefont {S.}~\bibnamefont
  {Klimenko}}, \bibinfo {author} {\bibfnamefont {I.}~\bibnamefont {Yakushin}},
  \bibinfo {author} {\bibfnamefont {A.}~\bibnamefont {Mercer}}, \ and\ \bibinfo
  {author} {\bibfnamefont {G.}~\bibnamefont {Mitselmakher}},\ }\href@noop {}
  {\bibfield  {journal} {\bibinfo  {journal} {Classical and Quantum Gravity}\
  }\textbf {\bibinfo {volume} {25}},\ \bibinfo {pages} {114029} (\bibinfo
  {year} {2008})}\BibitemShut {NoStop}%
\bibitem [{\citenamefont {Sutton}\ \emph {et~al.}(2010)\citenamefont {Sutton},
  \citenamefont {Jones}, \citenamefont {Chatterji}, \citenamefont {Kalmus},
  \citenamefont {Leonor}, \citenamefont {Poprocki}, \citenamefont {Rollins},
  \citenamefont {Searle}, \citenamefont {Stein}, \citenamefont {Tinto} \emph
  {et~al.}}]{sutton2010x}%
  \BibitemOpen
  \bibfield  {author} {\bibinfo {author} {\bibfnamefont {P.~J.}\ \bibnamefont
  {Sutton}}, \bibinfo {author} {\bibfnamefont {G.}~\bibnamefont {Jones}},
  \bibinfo {author} {\bibfnamefont {S.}~\bibnamefont {Chatterji}}, \bibinfo
  {author} {\bibfnamefont {P.}~\bibnamefont {Kalmus}}, \bibinfo {author}
  {\bibfnamefont {I.}~\bibnamefont {Leonor}}, \bibinfo {author} {\bibfnamefont
  {S.}~\bibnamefont {Poprocki}}, \bibinfo {author} {\bibfnamefont
  {J.}~\bibnamefont {Rollins}}, \bibinfo {author} {\bibfnamefont
  {A.}~\bibnamefont {Searle}}, \bibinfo {author} {\bibfnamefont
  {L.}~\bibnamefont {Stein}}, \bibinfo {author} {\bibfnamefont
  {M.}~\bibnamefont {Tinto}},  \emph {et~al.},\ }\href@noop {} {\bibfield
  {journal} {\bibinfo  {journal} {New Journal of Physics}\ }\textbf {\bibinfo
  {volume} {12}},\ \bibinfo {pages} {053034} (\bibinfo {year}
  {2010})}\BibitemShut {NoStop}%
\bibitem [{\citenamefont {Harry}\ and\ \citenamefont
  {Fairhurst}(2011)}]{harry2011targeted}%
  \BibitemOpen
  \bibfield  {author} {\bibinfo {author} {\bibfnamefont {I.~W.}\ \bibnamefont
  {Harry}}\ and\ \bibinfo {author} {\bibfnamefont {S.}~\bibnamefont
  {Fairhurst}},\ }\href@noop {} {\bibfield  {journal} {\bibinfo  {journal}
  {Physical Review D}\ }\textbf {\bibinfo {volume} {83}},\ \bibinfo {pages}
  {084002} (\bibinfo {year} {2011})}\BibitemShut {NoStop}%
\bibitem [{\citenamefont {Williamson}\ \emph {et~al.}(2014)\citenamefont
  {Williamson}, \citenamefont {Biwer}, \citenamefont {Fairhurst}, \citenamefont
  {Harry}, \citenamefont {Macdonald}, \citenamefont {Macleod},\ and\
  \citenamefont {Predoi}}]{williamson2014improved}%
  \BibitemOpen
  \bibfield  {author} {\bibinfo {author} {\bibfnamefont {A.~R.}\ \bibnamefont
  {Williamson}}, \bibinfo {author} {\bibfnamefont {C.}~\bibnamefont {Biwer}},
  \bibinfo {author} {\bibfnamefont {S.}~\bibnamefont {Fairhurst}}, \bibinfo
  {author} {\bibfnamefont {I.}~\bibnamefont {Harry}}, \bibinfo {author}
  {\bibfnamefont {E.}~\bibnamefont {Macdonald}}, \bibinfo {author}
  {\bibfnamefont {D.}~\bibnamefont {Macleod}}, \ and\ \bibinfo {author}
  {\bibfnamefont {V.}~\bibnamefont {Predoi}},\ }\href@noop {} {\bibfield
  {journal} {\bibinfo  {journal} {Physical Review D}\ }\textbf {\bibinfo
  {volume} {90}},\ \bibinfo {pages} {122004} (\bibinfo {year}
  {2014})}\BibitemShut {NoStop}%
\bibitem [{\citenamefont {Babak}\ \emph {et~al.}(2013)\citenamefont {Babak},
  \citenamefont {Biswas}, \citenamefont {Brady}, \citenamefont {Brown},
  \citenamefont {Cannon}, \citenamefont {Capano}, \citenamefont {Clayton},
  \citenamefont {Cokelaer}, \citenamefont {Creighton}, \citenamefont {Dent}
  \emph {et~al.}}]{babak2013searching}%
  \BibitemOpen
  \bibfield  {author} {\bibinfo {author} {\bibfnamefont {S.}~\bibnamefont
  {Babak}}, \bibinfo {author} {\bibfnamefont {R.}~\bibnamefont {Biswas}},
  \bibinfo {author} {\bibfnamefont {P.}~\bibnamefont {Brady}}, \bibinfo
  {author} {\bibfnamefont {D.~A.}\ \bibnamefont {Brown}}, \bibinfo {author}
  {\bibfnamefont {K.}~\bibnamefont {Cannon}}, \bibinfo {author} {\bibfnamefont
  {C.~D.}\ \bibnamefont {Capano}}, \bibinfo {author} {\bibfnamefont {J.~H.}\
  \bibnamefont {Clayton}}, \bibinfo {author} {\bibfnamefont {T.}~\bibnamefont
  {Cokelaer}}, \bibinfo {author} {\bibfnamefont {J.~D.}\ \bibnamefont
  {Creighton}}, \bibinfo {author} {\bibfnamefont {T.}~\bibnamefont {Dent}},
  \emph {et~al.},\ }\href@noop {} {\bibfield  {journal} {\bibinfo  {journal}
  {Physical Review D}\ }\textbf {\bibinfo {volume} {87}},\ \bibinfo {pages}
  {024033} (\bibinfo {year} {2013})}\BibitemShut {NoStop}%
\bibitem [{\citenamefont {Usman}\ \emph {et~al.}(2016)\citenamefont {Usman},
  \citenamefont {Nitz}, \citenamefont {Harry}, \citenamefont {Biwer},
  \citenamefont {Brown}, \citenamefont {Cabero}, \citenamefont {Capano},
  \citenamefont {Dal~Canton}, \citenamefont {Dent}, \citenamefont {Fairhurst}
  \emph {et~al.}}]{usman2016pycbc}%
  \BibitemOpen
  \bibfield  {author} {\bibinfo {author} {\bibfnamefont {S.~A.}\ \bibnamefont
  {Usman}}, \bibinfo {author} {\bibfnamefont {A.~H.}\ \bibnamefont {Nitz}},
  \bibinfo {author} {\bibfnamefont {I.~W.}\ \bibnamefont {Harry}}, \bibinfo
  {author} {\bibfnamefont {C.~M.}\ \bibnamefont {Biwer}}, \bibinfo {author}
  {\bibfnamefont {D.~A.}\ \bibnamefont {Brown}}, \bibinfo {author}
  {\bibfnamefont {M.}~\bibnamefont {Cabero}}, \bibinfo {author} {\bibfnamefont
  {C.~D.}\ \bibnamefont {Capano}}, \bibinfo {author} {\bibfnamefont
  {T.}~\bibnamefont {Dal~Canton}}, \bibinfo {author} {\bibfnamefont
  {T.}~\bibnamefont {Dent}}, \bibinfo {author} {\bibfnamefont {S.}~\bibnamefont
  {Fairhurst}},  \emph {et~al.},\ }\href@noop {} {\bibfield  {journal}
  {\bibinfo  {journal} {Classical and Quantum Gravity}\ }\textbf {\bibinfo
  {volume} {33}},\ \bibinfo {pages} {215004} (\bibinfo {year}
  {2016})}\BibitemShut {NoStop}%
\bibitem [{\citenamefont {Messick}\ \emph {et~al.}(2017)\citenamefont
  {Messick}, \citenamefont {Blackburn}, \citenamefont {Brady}, \citenamefont
  {Brockill}, \citenamefont {Cannon}, \citenamefont {Cariou}, \citenamefont
  {Caudill}, \citenamefont {Chamberlin}, \citenamefont {Creighton},
  \citenamefont {Everett} \emph {et~al.}}]{messick2017analysis}%
  \BibitemOpen
  \bibfield  {author} {\bibinfo {author} {\bibfnamefont {C.}~\bibnamefont
  {Messick}}, \bibinfo {author} {\bibfnamefont {K.}~\bibnamefont {Blackburn}},
  \bibinfo {author} {\bibfnamefont {P.}~\bibnamefont {Brady}}, \bibinfo
  {author} {\bibfnamefont {P.}~\bibnamefont {Brockill}}, \bibinfo {author}
  {\bibfnamefont {K.}~\bibnamefont {Cannon}}, \bibinfo {author} {\bibfnamefont
  {R.}~\bibnamefont {Cariou}}, \bibinfo {author} {\bibfnamefont
  {S.}~\bibnamefont {Caudill}}, \bibinfo {author} {\bibfnamefont {S.~J.}\
  \bibnamefont {Chamberlin}}, \bibinfo {author} {\bibfnamefont {J.~D.}\
  \bibnamefont {Creighton}}, \bibinfo {author} {\bibfnamefont {R.}~\bibnamefont
  {Everett}},  \emph {et~al.},\ }\href@noop {} {\bibfield  {journal} {\bibinfo
  {journal} {Physical Review D}\ }\textbf {\bibinfo {volume} {95}},\ \bibinfo
  {pages} {042001} (\bibinfo {year} {2017})}\BibitemShut {NoStop}%
\bibitem [{\citenamefont {Cannon}(2008)}]{cannon2008bayesian}%
  \BibitemOpen
  \bibfield  {author} {\bibinfo {author} {\bibfnamefont {K.~C.}\ \bibnamefont
  {Cannon}},\ }\href@noop {} {\bibfield  {journal} {\bibinfo  {journal}
  {Classical and Quantum Gravity}\ }\textbf {\bibinfo {volume} {25}},\ \bibinfo
  {pages} {105024} (\bibinfo {year} {2008})}\BibitemShut {NoStop}%
\bibitem [{\citenamefont {{Cannon}}\ \emph {et~al.}(2015)\citenamefont
  {{Cannon}}, \citenamefont {{Hanna}},\ and\ \citenamefont
  {{Peoples}}}]{2015likelihoodratio}%
  \BibitemOpen
  \bibfield  {author} {\bibinfo {author} {\bibfnamefont {K.}~\bibnamefont
  {{Cannon}}}, \bibinfo {author} {\bibfnamefont {C.}~\bibnamefont {{Hanna}}}, \
  and\ \bibinfo {author} {\bibfnamefont {J.}~\bibnamefont {{Peoples}}},\
  }\href@noop {} {\bibfield  {journal} {\bibinfo  {journal} {ArXiv e-prints}\ }
  (\bibinfo {year} {2015})},\ \Eprint {http://arxiv.org/abs/1504.04632}
  {arXiv:1504.04632 [astro-ph.IM]} \BibitemShut {NoStop}%
\bibitem [{\citenamefont {Nitz}\ \emph {et~al.}(2017)\citenamefont {Nitz},
  \citenamefont {Dent}, \citenamefont {Dal~Canton}, \citenamefont {Fairhurst},\
  and\ \citenamefont {Brown}}]{nitz2017detecting}%
  \BibitemOpen
  \bibfield  {author} {\bibinfo {author} {\bibfnamefont {A.~H.}\ \bibnamefont
  {Nitz}}, \bibinfo {author} {\bibfnamefont {T.}~\bibnamefont {Dent}}, \bibinfo
  {author} {\bibfnamefont {T.}~\bibnamefont {Dal~Canton}}, \bibinfo {author}
  {\bibfnamefont {S.}~\bibnamefont {Fairhurst}}, \ and\ \bibinfo {author}
  {\bibfnamefont {D.~A.}\ \bibnamefont {Brown}},\ }\href@noop {} {\bibfield
  {journal} {\bibinfo  {journal} {The Astrophysical Journal}\ }\textbf
  {\bibinfo {volume} {849}},\ \bibinfo {pages} {118} (\bibinfo {year}
  {2017})}\BibitemShut {NoStop}%
\bibitem [{\citenamefont {{Sachdev}}\ \emph {et~al.}()\citenamefont {{Sachdev}}
  \emph {et~al.}}]{2018sachdev}%
  \BibitemOpen
  \bibfield  {author} {\bibinfo {author} {\bibfnamefont {S.}~\bibnamefont
  {{Sachdev}}} \emph {et~al.},\ }\href@noop {} {\bibinfo  {journal} {{\it in
  prep.}}\ }\BibitemShut {NoStop}%
\bibitem [{\citenamefont {Abbott}\ \emph
  {et~al.}(2016{\natexlab{b}})\citenamefont {Abbott}, \citenamefont {Abbott},
  \citenamefont {Abbott}, \citenamefont {Abernathy}, \citenamefont {Acernese},
  \citenamefont {Ackley}, \citenamefont {Adams}, \citenamefont {Adams},
  \citenamefont {Addesso}, \citenamefont {Adhikari} \emph
  {et~al.}}]{abbott2016binary}%
  \BibitemOpen
\bibfield  {journal} {  }\bibfield  {author} {\bibinfo {author} {\bibfnamefont
  {B.}~\bibnamefont {Abbott}}, \bibinfo {author} {\bibfnamefont
  {R.}~\bibnamefont {Abbott}}, \bibinfo {author} {\bibfnamefont
  {T.}~\bibnamefont {Abbott}}, \bibinfo {author} {\bibfnamefont
  {M.}~\bibnamefont {Abernathy}}, \bibinfo {author} {\bibfnamefont
  {F.}~\bibnamefont {Acernese}}, \bibinfo {author} {\bibfnamefont
  {K.}~\bibnamefont {Ackley}}, \bibinfo {author} {\bibfnamefont
  {C.}~\bibnamefont {Adams}}, \bibinfo {author} {\bibfnamefont
  {T.}~\bibnamefont {Adams}}, \bibinfo {author} {\bibfnamefont
  {P.}~\bibnamefont {Addesso}}, \bibinfo {author} {\bibfnamefont
  {R.}~\bibnamefont {Adhikari}},  \emph {et~al.},\ }\href@noop {} {\bibfield
  {journal} {\bibinfo  {journal} {Physical Review X}\ }\textbf {\bibinfo
  {volume} {6}},\ \bibinfo {pages} {041015} (\bibinfo {year}
  {2016}{\natexlab{b}})}\BibitemShut {NoStop}%
\bibitem [{\citenamefont {Abbott}\ \emph
  {et~al.}(2016{\natexlab{c}})\citenamefont {Abbott}, \citenamefont {Abbott},
  \citenamefont {Abbott}, \citenamefont {Abernathy}, \citenamefont {Acernese},
  \citenamefont {Ackley}, \citenamefont {Adams}, \citenamefont {Adams},
  \citenamefont {Addesso}, \citenamefont {Adhikari} \emph
  {et~al.}}]{abbott2016gw151226}%
  \BibitemOpen
  \bibfield  {author} {\bibinfo {author} {\bibfnamefont {B.~P.}\ \bibnamefont
  {Abbott}}, \bibinfo {author} {\bibfnamefont {R.}~\bibnamefont {Abbott}},
  \bibinfo {author} {\bibfnamefont {T.}~\bibnamefont {Abbott}}, \bibinfo
  {author} {\bibfnamefont {M.}~\bibnamefont {Abernathy}}, \bibinfo {author}
  {\bibfnamefont {F.}~\bibnamefont {Acernese}}, \bibinfo {author}
  {\bibfnamefont {K.}~\bibnamefont {Ackley}}, \bibinfo {author} {\bibfnamefont
  {C.}~\bibnamefont {Adams}}, \bibinfo {author} {\bibfnamefont
  {T.}~\bibnamefont {Adams}}, \bibinfo {author} {\bibfnamefont
  {P.}~\bibnamefont {Addesso}}, \bibinfo {author} {\bibfnamefont
  {R.}~\bibnamefont {Adhikari}},  \emph {et~al.},\ }\href@noop {} {\bibfield
  {journal} {\bibinfo  {journal} {Physical review letters}\ }\textbf {\bibinfo
  {volume} {116}},\ \bibinfo {pages} {241103} (\bibinfo {year}
  {2016}{\natexlab{c}})}\BibitemShut {NoStop}%
\bibitem [{\citenamefont {Abbott}\ \emph
  {et~al.}(2016{\natexlab{d}})\citenamefont {Abbott}, \citenamefont {Abbott},
  \citenamefont {Abbott}, \citenamefont {Abernathy}, \citenamefont {Acernese},
  \citenamefont {Ackley}, \citenamefont {Adams}, \citenamefont {Adams},
  \citenamefont {Addesso}, \citenamefont {Adhikari} \emph
  {et~al.}}]{abbott2016gw150914}%
  \BibitemOpen
  \bibfield  {author} {\bibinfo {author} {\bibfnamefont {B.~P.}\ \bibnamefont
  {Abbott}}, \bibinfo {author} {\bibfnamefont {R.}~\bibnamefont {Abbott}},
  \bibinfo {author} {\bibfnamefont {T.}~\bibnamefont {Abbott}}, \bibinfo
  {author} {\bibfnamefont {M.}~\bibnamefont {Abernathy}}, \bibinfo {author}
  {\bibfnamefont {F.}~\bibnamefont {Acernese}}, \bibinfo {author}
  {\bibfnamefont {K.}~\bibnamefont {Ackley}}, \bibinfo {author} {\bibfnamefont
  {C.}~\bibnamefont {Adams}}, \bibinfo {author} {\bibfnamefont
  {T.}~\bibnamefont {Adams}}, \bibinfo {author} {\bibfnamefont
  {P.}~\bibnamefont {Addesso}}, \bibinfo {author} {\bibfnamefont
  {R.}~\bibnamefont {Adhikari}},  \emph {et~al.},\ }\href@noop {} {\bibfield
  {journal} {\bibinfo  {journal} {Physical Review D}\ }\textbf {\bibinfo
  {volume} {93}},\ \bibinfo {pages} {122003} (\bibinfo {year}
  {2016}{\natexlab{d}})}\BibitemShut {NoStop}%
\bibitem [{\citenamefont {{LIGO Scientific Collaboration and Virgo
  Collaboration}}(2018{\natexlab{a}})}]{ligo2018gstlal}%
  \BibitemOpen
  \bibfield  {author} {\bibinfo {author} {\bibnamefont {{LIGO Scientific
  Collaboration and Virgo Collaboration}}},\ }\href@noop {} {\enquote {\bibinfo
  {title} {Gstlal},}\ } (\bibinfo {year} {2018}{\natexlab{a}})\BibitemShut
  {NoStop}%
\bibitem [{\citenamefont {Cannon}\ \emph {et~al.}(2012)\citenamefont {Cannon},
  \citenamefont {Cariou}, \citenamefont {Chapman}, \citenamefont
  {Crispin-Ortuzar}, \citenamefont {Fotopoulos}, \citenamefont {Frei},
  \citenamefont {Hanna}, \citenamefont {Kara}, \citenamefont {Keppel},
  \citenamefont {Liao} \emph {et~al.}}]{cannon2012toward}%
  \BibitemOpen
  \bibfield  {author} {\bibinfo {author} {\bibfnamefont {K.}~\bibnamefont
  {Cannon}}, \bibinfo {author} {\bibfnamefont {R.}~\bibnamefont {Cariou}},
  \bibinfo {author} {\bibfnamefont {A.}~\bibnamefont {Chapman}}, \bibinfo
  {author} {\bibfnamefont {M.}~\bibnamefont {Crispin-Ortuzar}}, \bibinfo
  {author} {\bibfnamefont {N.}~\bibnamefont {Fotopoulos}}, \bibinfo {author}
  {\bibfnamefont {M.}~\bibnamefont {Frei}}, \bibinfo {author} {\bibfnamefont
  {C.}~\bibnamefont {Hanna}}, \bibinfo {author} {\bibfnamefont
  {E.}~\bibnamefont {Kara}}, \bibinfo {author} {\bibfnamefont {D.}~\bibnamefont
  {Keppel}}, \bibinfo {author} {\bibfnamefont {L.}~\bibnamefont {Liao}},  \emph
  {et~al.},\ }\href@noop {} {\bibfield  {journal} {\bibinfo  {journal} {The
  Astrophysical Journal}\ }\textbf {\bibinfo {volume} {748}},\ \bibinfo {pages}
  {136} (\bibinfo {year} {2012})}\BibitemShut {NoStop}%
\bibitem [{\citenamefont {Privitera}\ \emph {et~al.}(2014)\citenamefont
  {Privitera}, \citenamefont {Mohapatra}, \citenamefont {Ajith}, \citenamefont
  {Cannon}, \citenamefont {Fotopoulos}, \citenamefont {Frei}, \citenamefont
  {Hanna}, \citenamefont {Weinstein},\ and\ \citenamefont
  {Whelan}}]{privitera2014improving}%
  \BibitemOpen
  \bibfield  {author} {\bibinfo {author} {\bibfnamefont {S.}~\bibnamefont
  {Privitera}}, \bibinfo {author} {\bibfnamefont {S.~R.}\ \bibnamefont
  {Mohapatra}}, \bibinfo {author} {\bibfnamefont {P.}~\bibnamefont {Ajith}},
  \bibinfo {author} {\bibfnamefont {K.}~\bibnamefont {Cannon}}, \bibinfo
  {author} {\bibfnamefont {N.}~\bibnamefont {Fotopoulos}}, \bibinfo {author}
  {\bibfnamefont {M.~A.}\ \bibnamefont {Frei}}, \bibinfo {author}
  {\bibfnamefont {C.}~\bibnamefont {Hanna}}, \bibinfo {author} {\bibfnamefont
  {A.~J.}\ \bibnamefont {Weinstein}}, \ and\ \bibinfo {author} {\bibfnamefont
  {J.~T.}\ \bibnamefont {Whelan}},\ }\href@noop {} {\bibfield  {journal}
  {\bibinfo  {journal} {Physical Review D}\ }\textbf {\bibinfo {volume} {89}},\
  \bibinfo {pages} {024003} (\bibinfo {year} {2014})}\BibitemShut {NoStop}%
\bibitem [{\citenamefont {Allen}\ \emph {et~al.}(2012)\citenamefont {Allen},
  \citenamefont {Anderson}, \citenamefont {Brady}, \citenamefont {Brown},\ and\
  \citenamefont {Creighton}}]{allen2012findchirp}%
  \BibitemOpen
  \bibfield  {author} {\bibinfo {author} {\bibfnamefont {B.}~\bibnamefont
  {Allen}}, \bibinfo {author} {\bibfnamefont {W.~G.}\ \bibnamefont {Anderson}},
  \bibinfo {author} {\bibfnamefont {P.~R.}\ \bibnamefont {Brady}}, \bibinfo
  {author} {\bibfnamefont {D.~A.}\ \bibnamefont {Brown}}, \ and\ \bibinfo
  {author} {\bibfnamefont {J.~D.}\ \bibnamefont {Creighton}},\ }\href@noop {}
  {\bibfield  {journal} {\bibinfo  {journal} {Physical Review D}\ }\textbf
  {\bibinfo {volume} {85}},\ \bibinfo {pages} {122006} (\bibinfo {year}
  {2012})}\BibitemShut {NoStop}%
\bibitem [{\citenamefont {Schutz}(2011)}]{schutz2011networks}%
  \BibitemOpen
  \bibfield  {author} {\bibinfo {author} {\bibfnamefont {B.~F.}\ \bibnamefont
  {Schutz}},\ }\href@noop {} {\bibfield  {journal} {\bibinfo  {journal}
  {Classical and Quantum Gravity}\ }\textbf {\bibinfo {volume} {28}},\ \bibinfo
  {pages} {125023} (\bibinfo {year} {2011})}\BibitemShut {NoStop}%
\bibitem [{\citenamefont {Fairhurst}(2009)}]{fairhurst2009}%
  \BibitemOpen
  \bibfield  {author} {\bibinfo {author} {\bibfnamefont {S.}~\bibnamefont
  {Fairhurst}},\ }\href@noop {} {\bibfield  {journal} {\bibinfo  {journal} {New
  Journal of Physics}\ }\textbf {\bibinfo {volume} {11}},\ \bibinfo {pages}
  {123006} (\bibinfo {year} {2009})}\BibitemShut {NoStop}%
\bibitem [{\citenamefont {Golub}\ and\ \citenamefont
  {Van~Loan}(2012)}]{golub2012}%
  \BibitemOpen
  \bibfield  {author} {\bibinfo {author} {\bibfnamefont {G.~H.}\ \bibnamefont
  {Golub}}\ and\ \bibinfo {author} {\bibfnamefont {C.~F.}\ \bibnamefont
  {Van~Loan}},\ }\href@noop {} {\emph {\bibinfo {title} {Matrix
  computations}}},\ Vol.~\bibinfo {volume} {3}\ (\bibinfo  {publisher} {JHU
  press},\ \bibinfo {year} {2012})\BibitemShut {NoStop}%
\bibitem [{\citenamefont {Jones}\ \emph {et~al.}(2014)\citenamefont {Jones},
  \citenamefont {Oliphant},\ and\ \citenamefont {Peterson}}]{jones2014scipy}%
  \BibitemOpen
  \bibfield  {author} {\bibinfo {author} {\bibfnamefont {E.}~\bibnamefont
  {Jones}}, \bibinfo {author} {\bibfnamefont {T.}~\bibnamefont {Oliphant}}, \
  and\ \bibinfo {author} {\bibfnamefont {P.}~\bibnamefont {Peterson}},\
  }\href@noop {} {\  (\bibinfo {year} {2014})}\BibitemShut {NoStop}%
\bibitem [{\citenamefont {{LIGO Scientific Collaboration and Virgo
  Collaboration}}(2018{\natexlab{b}})}]{ligo2018lalsuite}%
  \BibitemOpen
  \bibfield  {author} {\bibinfo {author} {\bibnamefont {{LIGO Scientific
  Collaboration and Virgo Collaboration}}},\ }\href@noop {} {\enquote {\bibinfo
  {title} {Lalsuite: Lsc algorithm library suite},}\ } (\bibinfo {year}
  {2018}{\natexlab{b}})\BibitemShut {NoStop}%
\bibitem [{\citenamefont {Scientific}\ \emph {et~al.}(2017)\citenamefont
  {Scientific}, \citenamefont {Abbott}, \citenamefont {Abbott}, \citenamefont
  {Abbott}, \citenamefont {Acernese}, \citenamefont {Ackley}, \citenamefont
  {Adams}, \citenamefont {Adams}, \citenamefont {Addesso}, \citenamefont
  {Adhikari} \emph {et~al.}}]{scientific2017gw170104}%
  \BibitemOpen
  \bibfield  {author} {\bibinfo {author} {\bibfnamefont {L.}~\bibnamefont
  {Scientific}}, \bibinfo {author} {\bibfnamefont {B.}~\bibnamefont {Abbott}},
  \bibinfo {author} {\bibfnamefont {R.}~\bibnamefont {Abbott}}, \bibinfo
  {author} {\bibfnamefont {T.}~\bibnamefont {Abbott}}, \bibinfo {author}
  {\bibfnamefont {F.}~\bibnamefont {Acernese}}, \bibinfo {author}
  {\bibfnamefont {K.}~\bibnamefont {Ackley}}, \bibinfo {author} {\bibfnamefont
  {C.}~\bibnamefont {Adams}}, \bibinfo {author} {\bibfnamefont
  {T.}~\bibnamefont {Adams}}, \bibinfo {author} {\bibfnamefont
  {P.}~\bibnamefont {Addesso}}, \bibinfo {author} {\bibfnamefont
  {R.}~\bibnamefont {Adhikari}},  \emph {et~al.},\ }\href@noop {} {\bibfield
  {journal} {\bibinfo  {journal} {Physical Review Letters}\ }\textbf {\bibinfo
  {volume} {118}},\ \bibinfo {pages} {221101} (\bibinfo {year}
  {2017})}\BibitemShut {NoStop}%
\bibitem [{\citenamefont {Abbott}\ \emph
  {et~al.}(2017{\natexlab{a}})\citenamefont {Abbott}, \citenamefont {Abbott},
  \citenamefont {Abbott}, \citenamefont {Acernese}, \citenamefont {Ackley},
  \citenamefont {Adams}, \citenamefont {Adams}, \citenamefont {Addesso},
  \citenamefont {Adhikari}, \citenamefont {Adya} \emph
  {et~al.}}]{abbott2017gw170608}%
  \BibitemOpen
  \bibfield  {author} {\bibinfo {author} {\bibfnamefont {B.~P.}\ \bibnamefont
  {Abbott}}, \bibinfo {author} {\bibfnamefont {R.}~\bibnamefont {Abbott}},
  \bibinfo {author} {\bibfnamefont {T.}~\bibnamefont {Abbott}}, \bibinfo
  {author} {\bibfnamefont {F.}~\bibnamefont {Acernese}}, \bibinfo {author}
  {\bibfnamefont {K.}~\bibnamefont {Ackley}}, \bibinfo {author} {\bibfnamefont
  {C.}~\bibnamefont {Adams}}, \bibinfo {author} {\bibfnamefont
  {T.}~\bibnamefont {Adams}}, \bibinfo {author} {\bibfnamefont
  {P.}~\bibnamefont {Addesso}}, \bibinfo {author} {\bibfnamefont
  {R.}~\bibnamefont {Adhikari}}, \bibinfo {author} {\bibfnamefont
  {V.}~\bibnamefont {Adya}},  \emph {et~al.},\ }\href@noop {} {\bibfield
  {journal} {\bibinfo  {journal} {The Astrophysical Journal Letters}\ }\textbf
  {\bibinfo {volume} {851}},\ \bibinfo {pages} {L35} (\bibinfo {year}
  {2017}{\natexlab{a}})}\BibitemShut {NoStop}%
\bibitem [{abb(2017)}]{abbott2017gw170814}%
  \BibitemOpen
  \href {\doibase 10.1103/PhysRevLett.119.141101} {\bibfield  {journal}
  {\bibinfo  {journal} {Physical Review Letters}\ }\textbf {\bibinfo {volume}
  {119}} (\bibinfo {year} {2017}),\ 10.1103/PhysRevLett.119.141101}\BibitemShut
  {NoStop}%
\bibitem [{\citenamefont {Abbott}\ \emph
  {et~al.}(2017{\natexlab{b}})\citenamefont {Abbott}, \citenamefont {Abbott},
  \citenamefont {Abbott}, \citenamefont {Acernese}, \citenamefont {Ackley},
  \citenamefont {Adams}, \citenamefont {Adams}, \citenamefont {Addesso},
  \citenamefont {Adhikari}, \citenamefont {Adya} \emph
  {et~al.}}]{abbott2017gw170817}%
  \BibitemOpen
  \bibfield  {author} {\bibinfo {author} {\bibfnamefont {B.~P.}\ \bibnamefont
  {Abbott}}, \bibinfo {author} {\bibfnamefont {R.}~\bibnamefont {Abbott}},
  \bibinfo {author} {\bibfnamefont {T.}~\bibnamefont {Abbott}}, \bibinfo
  {author} {\bibfnamefont {F.}~\bibnamefont {Acernese}}, \bibinfo {author}
  {\bibfnamefont {K.}~\bibnamefont {Ackley}}, \bibinfo {author} {\bibfnamefont
  {C.}~\bibnamefont {Adams}}, \bibinfo {author} {\bibfnamefont
  {T.}~\bibnamefont {Adams}}, \bibinfo {author} {\bibfnamefont
  {P.}~\bibnamefont {Addesso}}, \bibinfo {author} {\bibfnamefont
  {R.}~\bibnamefont {Adhikari}}, \bibinfo {author} {\bibfnamefont
  {V.}~\bibnamefont {Adya}},  \emph {et~al.},\ }\href@noop {} {\bibfield
  {journal} {\bibinfo  {journal} {Physical Review Letters}\ }\textbf {\bibinfo
  {volume} {119}},\ \bibinfo {pages} {161101} (\bibinfo {year}
  {2017}{\natexlab{b}})}\BibitemShut {NoStop}%
\end{thebibliography}%

\end{document}